\documentclass[aps,manuscript,overcite]{revtex}
\usepackage{graphics}
\input epsf
\tighten
\def\ba{\begin{eqnarray}}
\def\ea{\end{eqnarray}}
\def\be{\begin{equation}}
\def\ee{\end{equation}}
\def\({\left(}
\def\){\right)}
\def\[{\left[}
\def\]{\right]}

\newcommand{\labeq}[1] {\label{eq:#1}}

\def\gsim{ \lower .75ex \hbox{$\sim$} \llap{\raise .27ex \hbox{$>$}} }
\def\lsim{ \lower .75ex \hbox{$\sim$} \llap{\raise .27ex \hbox{$<$}} }

\def\t{\tau}
\begin{document}


\title{Density Perturbations in the
 Ekpyrotic Scenario}

\author{ Justin Khoury$^1$, Burt A. Ovrut$^2$, Paul J. Steinhardt$^1$
and Neil Turok $^{3}$}

\address{
$^1$ Joseph Henry Laboratories,
Princeton University,
Princeton, NJ 08544, USA \\
$^2$ Department of Physics, University of Pennsylvania,
Philadelphia, PA 19104-6396, USA\\
$^3$ DAMTP, CMS, Wilberforce Road, Cambridge, CB3 0WA, UK}

\maketitle

\begin{abstract}
We study the generation of density perturbations in
the ekpyrotic scenario for the early universe, including 
gravitational
backreaction. We expose 
interesting subtleties 
that apply to both inflationary
and ekpyrotic models. 
Our analysis includes a detailed proposal of how the perturbations
generated in a contracting phase may be matched across a `bounce'
to those in an expanding hot big bang phase.  For
the physical conditions relevant to the ekpyrotic scenario, we re-obtain
our earlier result of a nearly scale-invariant spectrum of energy
density perturbations. We find that the perturbation amplitude is typically 
small, as desired to match observation.

\end{abstract}

\baselineskip 24pt

We recently proposed a novel scenario for the
early Universe
in which the hot big bang is created by the collision
between two M-theory 
branes\cite{kost}\null.  The scenario assumes the Universe begins in an
almost static, nearly BPS initial state consisting of empty, flat,
parallel three-branes.  In the effective 4d theory,
the BPS state is homogeneous and has zero spatial
curvature.  
Due to non-perturbative effects, however, a tiny 
force attracts the branes to one
another.  As the branes come together, quantum fluctuations create
ripples in the brane surfaces that result in spatial variations in the 
time of collision.  Consequently, 
some regions heat up and begin
to cool before others, producing a spectrum of long wavelength 
density perturbations which can seed structure formation in the 
Universe.

We estimated the  perturbation spectrum using a 
`time delay' formalism\cite{GuthPi}\null, often used in simplified
treatments of inflationary models. 
In that context, spatial variations in the time
when inflation ends result in long wavelength 
density inhomogeneities.  We applied the same
formalism to variations in the time of collision
in the ekpyrotic scenario.  The 
equation for  fluctuations in the scalar field $\phi$
describing the inter-brane separation in the ekpyrotic model
is almost identical to that describing fluctuations in
the inflaton during slow-roll inflation. 
Consequently, a nearly scale-invariant spectrum of fluctuations is 
found.  The result is remarkable because it shows that the 
Harrison-Zel'dovich spectrum can be obtained without inflation in a
space-time which is very nearly static Minkowski space.

The time delay formalism is a crude approximation,
and only quantitatively 
accurate for a small class of 
inflationary potentials \cite{Wang}\null. Nevertheless, 
it often gives a good
estimate of the spectral index for the power spectrum
of perturbations. One of the goals of this paper is to investigate
whether the same statement is 
true for the ekpyrotic model.

In the case of the ekpyrotic model, there is the major complication that
the perturbations are produced when the effective 4d scale factor
is contracting. In order to have a viable scenario, a mechanism must
be found to reverse from contraction 
to expansion.  This issue has been addressed in
a recent paper we have written with N. Seiberg\cite{nonsing}\null, where
we argue that such a `bounce' may be allowed in the context of M-theory,
where it 
corresponds to a collision and rebound of the outer boundary
branes.  A matching rule linking the homogeneous background variables
of the contracting phase to those describing the 
expanding phase was suggested there.  Assuming
this proposal is valid, what remains is to apply and extend those ideas
to describe the evolution of perturbations through the moment of
reversal. 

A stimulus for the present work
was a paper by D. Lyth\cite{lyth0}, which calculated the growth
of the perturbation variable $\zeta$ (also commonly termed ${\cal R}$) 
representing the curvature perturbation of spatial slices which
are comoving with the matter.
Lyth correctly showed that $\zeta$ was not amplified in the 
contracting phase of the ekpyrotic universe. He claimed this implied 
that
when gravitational back-reaction was included, the spectrum
of density perturbations became strongly scale-dependent with 
negligible power on large scales, making the ekpyrotic scenario
incompatible with observations.  
His analysis employed a certain class of analytically solvable
models with exponential potentials, previously used to describe power
law inflation\cite{lyth1} and simply extended
to the situation of slow contraction relevant to
the ekpyrotic scenario. We repeat his analysis here, but also 
compute the perturbation in the Newtonian potential $\Phi$.
We show that in the contracting phase
gravitational back-reaction actually enhances rather
than suppresses long wavelength fluctuations, but 
these fluctuations show up purely in $\Phi$ and not in $\zeta$.
Therefore gravitational back-reaction does not spoil the 
ekpyrotic mechanism, at least in the contracting phase.

The  remaining issue 
regards the appropriate  
matching condition for tracking the perturbations 
across the bounce and into the expanding hot big bang.
Consistent with the arguments of Ref. 4, we seek to
identify variables which 
are non-singular at the bounce, both for the background and
perturbation variables.
We then match 
the amplitudes of the two linearly independent 
solutions for the perturbation variables across the bounce.
With our prescription, we find that the 
long wavelength perturbations developed in the contracting phase
do indeed survive to the expanding phase, provided there
is a change in the equation of state at the bounce,
such as occurs if a sub-dominant component of radiation is produced there.
Our final expression for the density perturbation spectrum agrees
well with the more naive time delay estimate.  

After communicating a preliminary version of this paper to 
Lyth and R. Brandenberger, 
Lyth prepared a second manuscript\cite{lyth2} proposing that
contraction be matched to expansion 
on a time-slice of fixed energy 
density\cite{lyth2}\null. With this procedure,
he argues that the curvature perturbation 
$\zeta$ is conserved across the bounce. But since $\zeta$ does
not acquire a scale invariant spectrum
in the contracting phase, he argues that 
$\zeta$ will not have such a spectrum in the expanding phase,
where it represents the 
amplitude of growing mode adiabatic perturbations.
Lyth's conclusion is that any growing mode density
perturbations developed 
in the contracting phase match perfectly onto 
pure decaying mode perturbations in the expanding
phase. Brandenberger and Finelli\cite{brand2}, and Hwang\cite{hwang}, 
have recently produced
preprints repeating this argument. In a note added, at the end of
this paper, we explain why we do not believe these
conclusions are valid for the ekpyrotic scenarios proposed
in Refs. 1 and 4.

Let us outline our approach to the 
matching problem.
We want to evolve  background and perturbation variables
according to the appropriate field equations, 
all the way to 
to zero scale factor in the four dimensional effective theory. 
We identify a complete set of variables which are
non-singular at the bounce, and match those non-singular 
variables across it.
This prescription automatically 
excludes the variables $\Phi$ and $\zeta$,
both of which diverge. More generally, geometrical
quantities such as the synchronous gauge comoving metric
perturbation $h_{ij}$ also diverge. Indeed the
meaning of the three-geometry is 
unclear at zero scale factor. 

Instead our approach is essentially algebraic rather than
geometrical. We focus on gauge invariant 
perturbation variables which are consistently
small at all times and match these across the bounce at $t=0$. 
We argue that this matching would give consistent results
for an infinite class of perturbation variables so defined.
Our prescription 
can 
only be fully justified by a satisfactory 
microscopic description of the relevant degrees of freedom. 
Nevertheless, if string theory shows that the 
scale factor can truly pass through zero and bounce, then tracking
perturbative gauge-invariant degrees of freedom  which remain small and finite
seems likely to be the right approach to matching fluctuations
across the bounce.

Our analysis shall be performed entirely within the context
of four dimensional effective field theory. This 
does not capture all the low energy degrees of freedom relevant to 
the five (or indeed eleven) dimensional brane-world. As in
Ref. 1, we shall assume those other degrees of freedom 
are frozen, or at least so slowly varying that their
inclusion would not substantially alter the result. 
We shall focus here on single moduli field $\phi$ which
determines the outer brane separation in a brane-world Universe. 
In Ref. 1 we considered a model in which the perturbations
are produced by the collision of a bulk brane with one of the boundary
branes.  As we pointed out there, if the 4d effective theory is
valid, the scale factor in that theory
must continue to contract
until the outer branes collide and bounce.  In Ref. 4
we suggested a simplified model in which there is only one collision
between the boundary branes and no bulk brane is needed. The perturbations
are produced as the outer branes approach one another.  For simplicity,
here we shall restrict ourselves to this two-brane scenario, in which 
the same scalar field $\phi$ is
responsible both for the development of the perturbations,
and for describing the 
the outer-brane collision and 
bounce. 
Generalizations to bulk-boundary collisions, as
in the original scenario of Ref. 1, are a straightforward
extension and will 
be briefly mentioned when appropriate.

The outline of this paper is as follows.
In Section I we 
discuss the properties of the inter-brane potential relevant to
the ekpyrotic scenario, an issue referred to again in the note added
at the end of this paper.
We then
review the application of the time delay formalism to
the ekpyrotic scenario, as described in Ref. 1, showing that
a scale invariant spectrum of fluctuations is naturally predicted.
 In Section II, we show that including gravitational
back-reaction has, in Newtonian gauge, only negligible
effects on the fluctuations acquired by the scalar field. 
Section III is devoted to a discussion of the role of the
curvature perturbation $\zeta$ (or ${\cal R}$) 
conventionally used in the
analysis of inflationary models, which is in fact insensitive
to the growing mode perturbation in the contracting phase of
the ekpyrotic model.
This is further elaborated in Section IV where we show that
$\zeta$ is canonically conjugate to the
direction of amplification, which is proportional to the variable 
$\Phi$. As a consequence of 
Liouville's theorem, $\zeta$ is `squeezed' as the perturbations develop
during the collapsing phase of the ekpyrotic scenario.

Section V addresses the key moment of reversal,
which is when, in the four dimensional effective description,  the
Universe `bounces' as the scale factor $a$ hits zero.  Since the
four dimensional spacetime geometry is singular there, a
four dimensional geometrical description
using the Einstein frame metric is inappropriate. 
However, our approach is to identify a complete set of 
dynamical variables which remain finite
at the bounce and to match them at the transition from contraction to 
expansion.  We ignore divergent quantities as unphysical, attributing their
bad behavior to a 
singular choice of field and metric variables. 
We used this approach in Ref.~\ref{nonsing} to describe the homogeneous
background evolution during the bounce.  Here we shall show
that there is also 
a set of gauge invariant linear perturbation variables which are 
well-behaved at the bounce.  
When the theory is formulated in these variables, 
there is a well defined matching prescription,
even in the four dimensional effective field theory, which
seems to yield physically sensible results. 

In order for the well-behaved variables to exist, we emphasize that
it is important to have two conditions: (1)~there must be a free scalar 
field (the modulus $\phi$ in our case) whose kinetic energy diverges
at the bounce and (2) the scalar potential $V(\phi)$ must be 
sufficiently non-singular at the bounce. The simplest case is where
$V(\phi)$ approaches
zero at the bounce, as suggested by the mapping from M-theory to
weakly coupled string theory\cite{nonsing}\null.
If these two conditions are fulfilled, then 
it is possible to follow the perturbations and match 
at $a =0$.

Our main finding, in Sections V and VI,
is that as long as radiation is produced or fields are excited
at the outer-brane collision so that there is a jump in the 
first or second time derivative of the 
equation of state parameter $w\equiv P/\rho$, 
the scale invariant perturbation spectrum developed during
the contracting phase propagates through the bounce
and into the final expanding Universe. Intriguingly,
the final density perturbation amplitude derived in
Section VI is naturally suppressed by a small numerical coefficient,
by quantities which are automatically small
in Horava-Witten theory, and 
by  factors involving the efficiency with which 
brane kinetic energy is converted into radiation.
We comment on how this suppression 
may naturally explain the small perturbation
amplitude $\sim 10^{-5}$ we see via observations of the
cosmic microwave sky in the universe today.

\section{Time Delay Neglecting Gravity Pre-Collision}

We  first derive  the ekpyrotic spectrum in a very naive 
`time delay' approach which totally
neglects cosmological expansion,  gravity and all
moduli other than $\phi$. We also ignore the crucial element
of reversal from
contraction to expansion, which will be a critical aspect of
the discussion in Sections V and VI. As we shall see, despite the
fact that the time delay argument ignores these very important
features, it nevertheless comes close to matching our final answer. 
Thus, as is the case in inflation, the time delay argument 
turns out to be a convenient heuristic even though it is 
neither physically rigorous nor numerically accurate.

We assume that at large positive $\phi$ 
the effective potential governing the
evolution of $\phi$ takes the form
\be
V= - V_0 e^{-c \phi},
\labeq{pot}
\ee
where $V_0$ and $c$ are positive constants.
The ekpyrotic scenario
starts at large positive $\phi$, in a state of nearly
zero energy, and rolls towards
negative values. 
As $\phi$ becomes increasingly negative and the branes come close together,
$V(\phi)$ must turn upwards and approach zero.  As discussed 
in Ref.~\ref{nonsing}, for example,
 this is the behavior
to be expected in M-theory
because the string coupling constant vanishes as the outer branes collide.
(A similar constraint applies in the bulk brane collision, as discussed
in Ref.~1.)

Generally, in brane world models,
the separation of the two boundary branes can be described 
by a canonically normalized, 
minimally coupled scalar field $\phi$ (the `radion'). 
As the brane separation $d$ goes to zero, one can ignore
the effect of the bulk `warp factor' and 
one obtains the 
Kaluza-Klein result in which 
$ d \sim e^{\sqrt{2/ 3} \phi/M_{Pl}} \rightarrow 0$
as $\phi \rightarrow -\infty$. (Here and below,
$M_{Pl} = (8\pi G)^{-{1\over 2}}$ is the four dimensional
reduced Planck mass.) In standard Kaluza-Klein theory,
the same dependence holds at large $\phi$, so the 
range of $\phi$ is $-\infty <\phi <\infty$. But 
the presence of a bulk warp factor alters this.
In many models, when the outer-brane distance 
$d$ tends to infinity, 
$\phi$ tends to a finite value, which may be taken to
be zero. For example, for  a pair of positive and negative tension
boundary branes 
with a bulk Anti-de Sitter space, the inter-brane distance
is $d=L$ ln(coth($-\phi/
\sqrt{6} M_{Pl}))$ where
$L$ is the AdS radius \cite{kost}\null.
A suitable ekpyrotic  potential 
would then behave as  $\sim -(-\phi)^N \sim -e^{(-N d/L)}$,
at small $\phi$, and we would be interested in starting the
system in a state such that $\phi \rightarrow 0$ at
large negative times.
In models with a bulk brane, such as those considered in 
Ref. 1, the 
location of the 
bulk brane would again be described by a scalar field, but this time
its range would be finite. 
The exponential model 
(\ref{eq:pot}) is useful, since
it is mathematically tractable. However, one should remember,
in general it should only be expected to apply over some finite range of
$\phi$.

We now review how the time delay formalism can be applied to our 
example.
The
classical solution is given by solving $\dot{\phi} = - \sqrt{-2 V}$.
For 
$V(\phi)$ given in (\ref{eq:pot}), we obtain
\begin{equation} \label{eq1}
-t = \int_{-\infty}^\phi { d\phi  \over \sqrt{2 V_0}} e^{c \phi/2}
   = {2\over  c \sqrt{2 V_0}} e^{c \phi/2}= \sqrt{2 \over -V_{,\phi \phi}}.
\end{equation}
where the time $t$ is large and negative at large $\phi$. We follow
the evolution to some finite, small negative $t$,
at which point the modes of interest are `frozen in'.
For small fluctuations, expanding in plane waves
$\delta \phi\equiv \sum_{\vec{k}} e^{i \vec{k}\cdot \vec{x}}
\delta \phi_{\vec{k}}(t)$, one finds
\begin{equation} \label{eq2}
\delta \ddot{\phi}_{\vec{k}}  =
- k^2 \delta \phi_{\vec{k}} - V_{,\phi \phi}  \delta \phi_{\vec{k}}
 =  - k^2 \delta \phi_{\vec{k}} +{2 \over t^2} \delta
 \phi_{\vec{k}}
\end{equation}
(using the solution  Eq.~(\ref{eq1}) above).
Curiously, even though the background here is nearly static,
this is the same equation as that
 governing  a massless field $\chi$
in a de Sitter background
(with
$\chi=\phi /a$, with $a$ the scale factor and $t$ 
conformal time). As in that case, starting from an initial
Minkowski vacuum one generates a long wavelength spectrum of
scale invariant fluctuations.

We assume that the quantum fluctuation $\delta \phi_{\vec{k}}$ starts
in the Minkowski vacuum, 
$\langle |\delta \phi_{\vec{k}}|^2 \rangle  \propto k^{-1}$,
so the amplitude of normalized modes with wavenumber
 $k$ is $\sim k^{-1/2}$.
Consider a mode where $ -k t_i \gg 1$,  where $t_i$ is the
initial time.
 The mode oscillates at fixed amplitude
$k^{-1/2}$ until $t$ increases to 
$t_1 \approx - k^{-1}$, when the destabilizing $2/t^2$ term in 
(\ref{eq2}) begins to dominate over the stabilizing $k^2$ term. 
One then has 
$ \delta \ddot{\phi}_{\vec{k}} \sim (2/t^2)\delta \phi_{\vec{k}}$,
with growing solution $\delta \phi_{\vec{k}}
\propto (-t)^{-1}$.
Hence, the final
amplitude for the fluctuation, as  $t$ approaches zero, is
$ \delta  \phi_{\vec{k}} \sim  k^{-1/2} (-t_1)/(-t) \sim k^{-3/2}(-t)^{-1}$.
The precise result for the long wavelength spectrum is
\be
\langle |\delta \phi_{\vec{k}}|^2\rangle = {1\over 2 k^3 t^2}, \qquad 
-kt <<1,
\labeq{noexp}
\ee
which is scale-invariant as claimed.

A naive estimate of the final density 
perturbation amplitude runs as follows.
The scalar field fluctuations give rise to a time delay in the moment
of collision. Neglecting gravitational effects,
the time delay between collisions relative to the mean
collision time is $-\delta \phi/\dot{\phi}_0$.
The net density perturbation after collision is then just
${\delta \rho \over \rho} \sim
- 4 H \delta t$
where $H$ is the Hubble parameter at the time of collision.
Despite the deficiencies of this time delay approach, this
answer is quite close to the result we shall eventually
derive, including gravitational back-reaction and matching
across the bounce, 
which we give in equation (\ref{eq:finalampf}) below.

The first step in improving the above treatment is
to 
include 
gravitational backreaction in 
the initial, contracting phase. 
For this
purpose, we focus on the
gravitational backreaction corrections to Eq.~(\ref{eq:noexp}).

\section{Including Gravitational Backreaction}

For exponential potentials, there exist analytic scaling solutions
to the Friedmann-Robertson-Walker
equations which at large negative $t$ approximate the assumed
initial conditions in the ekpyrotic setup, and which
allow an analytic treatment
\cite{lyth1}\null. 
For the exponential potential considered as an approximation over
some range of $\phi$ in the ekpyrotic scenario,
one has the background solution
\be
a(t)= (-t)^p, \qquad \phi_0(t)={2\over c} {\rm ln} (-M t),
\labeq{back}
\ee
where $a(t)$ is the scale factor, $t$ is proper FRW time, taking negative
values, and 
\be
p= {2\over c^2 M_{Pl}^2}, \qquad M^2= { V_0 \over M_{Pl}^2 p(1-3p)},
\labeq{params}
\ee
and  $M_{Pl} = (8 \pi G)^{-{1\over2}}$ is the
reduced Planck mass.
Some useful quantities are:
\be
H\equiv {\dot{a}\over a} = {p\over t}, \qquad 
\dot{\phi}_0= \sqrt{2p} {M_{Pl} \over t}, \qquad 
V=-p(1-3p) {M_{Pl}^2 \over t^2}, \qquad
V_{,\phi \phi} = -{2\over t^2} (1- 3p),
\labeq{quants}
\ee
where dot denotes $d/dt$.
In ekpyrotic models, we are interested in potentials in which
the evolution is very slow or, equivalently, for which 
$p <<1$. (Parenthetically,
note that the same solution, for $p>1$, describes a universe 
undergoing power law inflation). 
The conformal time is
\be
\tau= - \int_t^0 {dt \over a(t)} = - {(-t)^{1-p} \over (1-p)} = 
{t \over a (1-p)},
\labeq{conft}
\ee
running from $-\infty$ to $0$ as $t$ does, although as mentioned
we shall have to
alter the form of the potential as $t$ and $\tau$ approach zero,
to avoid its diverging to minus infinity.
Derivatives with respect to $t$ shall, as above, 
be denoted by dots and henceforth, derivatives
with respect to $\tau$ shall be denoted by primes.

Mathematically,
Newtonian gauge is  convenient for studying scalar field
perturbations
since the linearized equations reduce to a single second
order differential equation for the Newtonian potential 
$\Phi$ (for a derivation
see e.g. Ref. \ref{bgt}),
\be
\Phi'' +2 {A'\over A} \Phi' +k^2 \Phi +
2 \phi_0'\left({{\cal H} \over \phi'_0}\right)' \Phi=0,
\labeq{Phieq}
\ee
where $A\equiv (\dot{\phi}_0)^{-1}$, and ${\cal H}\equiv a'/a$. Here, and below,
the wavenumber dependence of all perturbation quantities is not
shown explicitly. For $\Phi$ one should read $\Phi_{\vec{k}}$ and 
so on.

We now eliminate the first
derivative term by setting $\Phi=u/A$, obtaining
\be
u''-{A''\over A} u +k^2 u +2 \phi_0'\left({H\over \dot{\phi}_0}\right)' u=0.
\labeq{neweq}
\ee
Substituting the above scaling solution, we obtain
\be
u''=-k^2 u + {p\over (1-p)^2 \tau^2} u,
\labeq{feq}
\ee
which is a form of Bessel's equation, with corresponding order $\nu={1\over 2} 
(1+p)/(1-p)$.
Note that our notation for 
the variable $u$, and for the 
variable $v$ introduced later, 
matches Mukhanov's original notation \cite{Mukh}\null.

The initial conditions are that the scalar field fluctuations should
be in the Minkowski vacuum state as $\t\rightarrow -\infty$: 
in Newtonian gauge,
two constraints determine $\Phi$ and $\dot{\Phi}$ in terms
of $\delta \phi$ and $\dot{\delta \phi}$:
\ba
\delta \phi &=&
{2 M_{Pl}^2 \over \dot \phi_0}
\left[ \dot \Phi  +H \Phi \right] ,\cr
\dot{\delta \phi} &=& - {2 M_{Pl}^2
\over \dot \phi _0 } \left[
 -{\ddot \phi _0\over \dot \phi _0}\dot \Phi
+\left\{ {k^2 \over a^2}+
\dot \phi _0 ~\partial _t\left( {H\over \dot \phi _0}\right)
\right\}  \Phi
\right] .
\labeq{phicons}
\ea
At large $-kt$, for $\delta \phi \sim e^{-i k \tau} /(a \sqrt{2k})$,
which is the incoming Minkowski vacuum, one finds
\be
\Phi \sim {i \sqrt{p} \over 2 M_{Pl} k^{3\over 2} t} e^{- i k \tau}
\labeq{phiin}
\ee
which vanishes at early times,  consistent with our
ekpyrotic initial condition that the  space is asymptotically
Minkowski in the past.

Neglecting  an irrelevant phase factor, the solution for $u$ is then
\be
u= (-k \tau)^{1\over 2} \sqrt{{\pi\over 2}} H_\nu^1(-k\tau )
 \cdot {\sqrt{p}
\over (2 k)^{3\over 2} M_{Pl}},
\labeq{fsol}
\ee
where $H_\nu^1$ is a Hankel function and $\nu$ was given above.
We follow this solution forward to conformal times at which
$-k\tau <<1$,
when the modes become frozen in. Converting back to $\Phi$,
and using the small argument expansion for the Hankel function,
again neglecting phase factors we find
\be
\Phi \sim {\sqrt{p} \over 2M_{Pl} } k^{-(1+\nu)} (-\tau)^{-2\nu}(1+O(p)).
\labeq{phisol}
\ee
Since we are interested in $p<<1$, we henceforth 
neglect corrections of order $p$ to the numerical
coefficient. But we keep the $p$ dependence in the 
scaling with wavelength and time for comparison with later results. 

Now we can convert back to the scalar field $\delta \phi$, using
the formulae (\ref{eq:phicons}) given above,
and  obtain:
\be
\delta \phi = {2 M_{Pl}^2 \over \dot{\phi}_0}
\left[\dot{\Phi}+H \Phi\right]
 \sim  2^{-1/2} k^{-(1+\nu)}
(-t)^{-(1+p)}.
\labeq{dps}
\ee
where we used $\Phi \propto (-\tau)^{-2 \nu} \propto (-t)^{-(1+p)}$.
The resulting power spectrum of fluctuations is:
\be
\langle |\delta \phi_{\vec{k}}|^2 \rangle \sim
\frac{1}{2}  k^{-2(1+\nu)}
(-t)^{-2(1+p)}
 =   \langle |\delta \phi_{\vec{k}}|^2\rangle^{(0)} { 1
\over k^{2p/(1-p)} (-t)^{2p}},
\labeq{dpsa}
\ee
where 
$\langle |\delta \phi_{\vec{k}}|^2\rangle^{(0)} $ is 
the result obtained in Eq.~(\ref{eq:noexp})
ignoring gravitational backreaction.
Recalling that
the regime of interest for the ekpyrotic scenario is small $p$,
we see that we obtain the same answer as in Eq.~(\ref{eq:noexp})
up to small corrections.
 The backreaction produces 
 a slight reddening of the power
spectrum, presumably because long wavelength perturbations were
generated sooner and had longer to self-gravitate.
However, since in the ekpyrotic scenario we have 
$p<<1$,
these corrections to the spectral
index are actually smaller than corrections arising from
the nontrivial kinetic term for $\phi$, which were computed
in Ref. 1. The result shows that backreaction
and metric fluctuations have an insignificant effect on the density
perturbations developed during the initial, contracting
phase of the ekpyrotic universe.

\section{The curvature perturbation in the contracting phase}

Rather than solve for the evolution of the
Newtonian potential in  Eq.~(\ref{eq:Phieq}),
a common procedure used for 
inflationary models  is to track  the curvature perturbation on
spatial slices which are comoving with the matter.
As we shall discuss,
this variable is in fact insensitive to the growing mode
density perturbation in the contracting phase.

Following the notation of Mukhanov {\it et al.} \cite{Mukh},
we denote the 
curvature perturbation on comoving slices by
$\zeta$,
defined by
\be
\zeta \equiv
\frac{2}{3} \frac{ {\cal H}^{-1} \Phi' + \Phi}{1+w} + \Phi 
= {{\cal H} \Phi' +{\cal H}^2 \Phi \over 4 \pi G \phi_0'^2} +\Phi,
\ee
where ${\cal H}\equiv a'/a$ and
$w$ is the ratio of pressure to density in the background universe.
The variable $\zeta$
was introduced by Bardeen\cite{bardeen} in his classic paper,
in which it was termed $\phi_m$. It was employed in the context
of inflation by Bardeen, Steinhardt, and Turner \cite{BST}
and its use later elaborated upon by many authors
\cite{Mukh,Brand,lyth1}\null.
Mukhanov {\it et al.} \cite{Mukh}  re-express $\zeta$ as
$\zeta =   v/z$  (for spatially flat hypersurfaces)
and thereby derive the `$v$-equation':
\begin{equation} \labeq{ueq}
v'' =- k^2 v  + \frac{z''}{z} v
\end{equation}
where $z = a \phi_0'/{\cal H}$. For the 
power law solution given in equation (\ref{eq:quants}),
this reads:
\begin{equation} \label{ueqp}
v'' = - k^2 v  - {p(1-2p)\over (1-p)^2 \tau^2 } v.
\end{equation}
Since the last term is negative, it follows that there is no
classical instability in the variable $v$, a point emphasized
by Ref.~3.
(Note that Ref.~3 relabels
Mukhanov's $v$ variable as $u$; here we have kept to the original notation.)

However, this does not at all imply the absence of a growing
mode density perturbation:
the variable
$\Phi$ {\it does} exhibit an instability, as we have discussed above.
Given a solution $v$, the potential $\Phi$ can be obtained
by solving 
\begin{equation} \labeq{sol1}
k^2 \Phi = - 4 \pi G  {\phi_0'^2 \over {\cal H}} (v/z)'.
\end{equation}
The general solution to the $v$-equation to order $k^2$ is:
\be
v  =   C_1 z \left( 1- k^2 \int^{\tau}_{-\infty} 
 \frac{d \tau'}{z^2} \int^{\tau'}_{-\infty} 
 z^2
d \tau''\right)   
+ C_2 z \left( \int^{\tau}_{-\infty} 
 \frac{d \tau'}{z^2} -
k^2 \int^{\tau}_{-\infty} 
 \frac{d \tau'}{z^2} \int^{\tau'}_{-\infty} z^2
d \tau''\right) ,
\ee
where $C_1$ and $C_2$ are arbitrary constants.
Substituting into Eq.~(\ref{eq:sol1}), we obtain
\begin{equation}
\Phi =   \bar{C}_1 \Phi_0 \int^{\tau}_{-\infty} 
 z^2 d \tau' +\bar{C}_2
\Phi_0
\end{equation}
where $\Phi_0 \equiv a'/a^3$,  and $\bar{C}_0$ 
and $\bar{C}_2$ are
constants.
For the power law solutions studied here,
$z \propto a$.
For an expanding universe, 
the first
term  dominates the second as  $t \rightarrow \infty$. However,
for a contracting universe,
the second term is the proper growing mode
as $t \rightarrow 0$.    The solution is the same as that
found in Eq.~(\ref{eq:phisol}), from which we derived the
scale-invariant spectrum.

Various subtleties are worthy of further comment. First
note that $\Phi_0 \equiv a'/a^3$ is an {\it exact} solution
of the perturbation equation (\ref{eq:Phieq}) for 
$k=0$. In fact, for $k=0$, the perturbation $\Phi_0$ 
just represents a coordinate transformation of the the time
$t$, and is therefore unphysical.
However, for 
nonzero $k$ the solutions
tending towards $\Phi_0$ are not gauge modes and therefore
represent a real gravitational instability.

Second, note that 
$\zeta$ 
is insensitive to any component of
$\Phi$ tending towards $\Phi_0$, 
as may be seen from the expression
\begin{equation}
\zeta = \frac{2}{3 a^2 (1+w)} \left( \frac{\Phi}{a'/a^3} \right)'.
\labeq{zintp}
\end{equation}
 For inflationary models, tracking $\zeta$, as is
done in many analyses, is  useful
because the components that are projected 
out in this formula are harmless decaying modes. 
However, for
the ekpyrotic model, the growing mode tends to
$\Phi_0$. One can in fact recover the 
the growing mode solution 
from $\zeta$ or $v$, but special care has to be 
taken to do so, by keeping the next
order terms in $k^2$ as we have done above.

\section{Canonical Conjugates and `Squeezing'}

In this section 
we show that the
variables $u$ in Eq.~(\ref{eq:neweq})
and $v$ in Eq.~(\ref{eq:ueq})
at the heart of the analysis of the previous
sections are in
fact
canonically
conjugate. The importance of this is that in the $(p,q)$ phase 
space plane, the trajectories of the system are focused towards certain lines
- for example the line $p=q$ in the case of the upside-down
harmonic oscillator. These lines are then the classically amplified
directions in phase space, and from Liouville's theorem
the phase space density is necessarily
squeezed in the orthogonal directions. While the 
calculation of Reference 5 is indeed correct, it does not by itself indicate
the absence of a growing mode density perturbation, because 
the curvature perturbation which was computed is actually 
{\it orthogonal} to the direction in which fluctuations are 
amplified in the ekpyrotic setup. (We note that References
\ref{lyth2} and \ref{branden} agree with this conclusion).

Our starting point is the action for gravity plus a scalar field in
canonical (first order) form. The relevant formulae may be found
for example in Ref. \ref{gt}. The Newtonian potential $\Phi$ is written in terms
of $u$,  and $\zeta$ is written in
terms of $v$ as above. One finds the quadratic action for perturbations
reduces to
\be
{\cal S} = {2 M_{Pl}^2 \over p} \int d\tau \left( v F' + {\beta \over \tau} v F
-{1\over 2} F^2 - {k^2\over 2} v^2 \right),
\labeq{act}
\ee
where $\beta= p/(1-p)$,
$F\equiv k^2 u$, and we have written the Laplacian $\nabla^2$ as $- k^2$.
The sum over all Fourier modes is implicit.

The action (\ref{eq:act}) is in canonical form
$\int p q'-H(p,q)$, with $H$ the Hamiltonian. Therefore,
up to normalization, $v$ and $F$ are canonically conjugate.
The equations of motion are
\be
F=-\tau^\beta\left({v \over \tau^\beta}\right)' \qquad k^2 v = \tau^{-\beta}
\left({F \tau^\beta}\right)',
\labeq{eoms}
\ee
which are just the relations above: equation 
(\ref{eq:sol1}), which expresses 
$\Phi$ in terms of $\zeta$ and $\zeta'$,  and  equation 
(\ref{eq:zintp}),
which expresses 
$\zeta$ in terms of $\Phi$ and $\Phi'$. These imply the
equations of motion (\ref{eq:feq}) and (\ref{eq:ueq}) used in the
previous sections.

In the inflationary case, which corresponds
to $p>1$ in the above analysis, both the variable $v$ and its canonical 
conjugate momentum $F$ (or $u$) obey equations exhibiting a classical
instability: both are amplified as $\tau$ approaches zero. The
system starts near $v=u=0$, with  
only small quantum fluctuations. As time proceeds the 
system evolves away from the origin 
along a certain 
line, $u \propto v$ in phase space. By Liouville's theorem, 
the phase space density is `squeezed' in the direction 
orthogonal to
this line. 
But in the ekpyrotic case, $p<<1$, 
the $v$ (or $\zeta$) variable is instead stable and 
classical trajectories instead run out along the
$u$ (or $\Phi$) axis. The  $v$ (or $\zeta$) direction
must in consequence be `squeezed'. Thus there is no classical
growth in $\zeta$, consistent with the finding 
of Reference \ref{lyth0}.

\section{Reversal and A Nonsingular Matching Condition}

Having established that there is indeed a growing mode
scale-invariant spectrum of density perturbations in the
collapsing phase, we now turn to the more challenging question 
of whether these perturbations survive the reversal to 
expansion. We have recently addressed the issue of reversal
in a paper with N. Seiberg\cite{nonsing}, and we shall employ and extend the
analysis of that paper here.

As we have discussed, the variable $\zeta$ is
not amplified in the collapsing phase, since it is insensitive to
the growing mode perturbation. However,
$\zeta$ {\it does} yield 
the amplitude of the growing mode perturbation in the expanding 
phase. Therefore, if 
$\zeta$ was in fact the correct variable to
match at $\tau=0$, 
the growing mode perturbation which developed in the collapsing
phase would just match onto a pure decaying mode 
perturbation in the expanding phase.
This has indeed 
been suggested by Lyth in a follow-up paper
following our communication of the above calculations to him \cite{lyth2}\null.
The  new argument is essentially that perturbation evolution may 
reverse at collision.
That is, $\Phi$ began
negligibly small, grew to be very 
large at the bounce but then, after
reversal to expansion, $\Phi$ shrank precisely back to zero.

In this section, we wish to explain why matching $\zeta$ across the bounce
is not appropriate in the 
scenarios we consider. Instead, we claim, 
the scale invariant 
growing mode spectrum generated in the contracting phase
matches onto a linear combination of growing and decaying modes
at reversal. Consequently, the density perturbations
survive the reversal of the scale factor to seed structure formation
in the hot big bang. 

In order to analyze matching at the bounce, gauge invariant 
variables must be identified which are well-behaved 
as $a$ and $\tau$ approach zero.
We have already seen that the growing 
mode solution for $\Phi$ is proportional to  $a'/a^3$, which is 
divergent as $\tau$ tends to zero. 
 Likewise, as discussed below, in the situation
of interest, $\zeta$ is logarithmically
divergent. Since linear perturbation theory can only be valid when the
perturbation variables are small, it follows that
$\zeta$ is not a good matching variable. Nevertheless,
with careful treatment to exclude the logarithmic
divergence, 
$\zeta$ shall be very useful in the analysis 
after $\tau=0$, just because it's
`long wavelength' component is 
nearly constant in the expanding phase, and gives the amplitude
of the growing mode linear density perturbation.

As mentioned in the introduction, 
the potential $V(\phi)$ 
in (\ref{eq:pot}) cannot hold as $\phi \rightarrow -\infty$.
In the main example we treat here, we instead 
assume $V(\phi)$
bends upward so that $V(\phi) \rightarrow 0$
as $\phi \rightarrow -\infty$. In this case, just before
collision the system is completely described as 
a massless scalar field coupled to gravity,
in a collapsing Universe. As discussed in 
Ref. 4, one can in this case define regular variables, in
which the evolution through $a=0$ becomes well defined. 

For the perturbation analysis, it is very important 
that, as $a\rightarrow 0$, the Universe
is dominated by scalar field kinetic energy. 
In this situation, a good perturbation variable is 
the fractional energy density perturbation on
spatial slices which are comoving with the matter,
termed $\epsilon_m$ by Bardeen in
his discussion of gauge invariant cosmological perturbation theory
\cite{bardeen}. This variable (defined in his equation 
3.13)  is a linear combination
of the energy density perturbation $\delta T^0_0$, and
the velocity 
perturbation $T^0_i$, which is invariant under linearized coordinate
transformations. It equals the energy density perturbation
in any gauge in which the matter worldlines are orthogonal
to the $\tau=$constant hypersurfaces, and as Bardeen emphasized
it is the natural choice of perturbation amplitude from
the point of view of the matter.

The equations of motion for the matter, $T^\mu_{\nu;\mu}=0$,
lead to the following equations of motion, 
\ba
(\rho a^3 \epsilon_m)'&=&-(\rho+p)a^3 k v_s^{(0)}\cr
{v_s^{(0)}}'+{\cal H} v_s^{(0)} &=& k \Phi +{k \over (1+w)} (c_s^2 \epsilon_m
+w \eta), 
\labeq{neweeq}
\ea
where $v_s^{(0)}$ is the gauge-invariant velocity perturbation 
defined by Bardeen. Here we introduce $w=p/\rho$ parameterizing 
the equation of state for pressure $p$ and 
energy density $\rho$,
$c_s^2=dp/d\rho$ being the sound speed,
and $\eta$ being the `entropy perturbation', 
defined as the 
difference between the pressure perturbation and  that expected
from the density perturbation and the background pressure-density
relation. In most of the analysis below we shall assume
$\eta$ is
zero, which is to say that the density perturbations are `adiabatic'.
For the types of matter we consider (perfect fluids and scalar
fields), the anisotropic stress is zero, and 
Bardeen's potentials are related to our potential $\Phi$ by
$\Phi=\Phi_A=-\Phi_H$. With these simplifications, and for the
flat universe we consider,
equations (4.5) and
(4.8) of Bardeen's paper\cite{bardeen} reduce to (\ref{eq:neweeq})
above.

Remarkably,
scalar field kinetic energy
domination implies that
$\epsilon_m$  is actually finite at $\tau=0$.
This can be seen from the Einstein constraint equation, which reads
\be
\nabla^2 \Phi =  4 \pi G \rho a^2 \epsilon_m,
\labeq{poisson}
\ee
where $\rho$ is the total background energy density. For
scalar field kinetic domination,
$\rho \propto a^{-6}$ as $a$ tends to zero. But the
growing mode perturbation $\Phi \propto a'/a^3 \propto
a^{-4}$ and, from Eq. (\ref{eq:poisson}), $\epsilon_m$ is finite
as $a$ tends to zero.

Having identified a non-singular matching variable, we still
need to decide which time slice to match on. In the present situation,
this is unambiguously defined as time slice where $\phi$ tends to
$-\infty$, or, re-phrased in terms of the brane separation, the
time when 
$d \propto e^{\sqrt{2/ 3} \phi/M_{Pl}} \rightarrow 0$. 
If we assume that no radiation is present in the
contracting phase, so that the only energy density present is that
in $\phi$, then the surfaces of constant $\phi$ are also the comoving
surfaces (because the perturbation to the momentum is proportional to
$\delta \phi$), and $\epsilon_m$ is the fractional 
energy density perturbation evaluated upon these
surfaces. 

At the bounce, we assume that there is a change in the internal state
of the branes so that $V(\phi)$ switches off and the
boundary branes no longer attract. 
After the bounce, 
$\phi$ is therefore a massless, free field. 
We also assume some 
radiation is produced on the branes. 
Again the matching
surface is defined in terms of $\phi$, but with 
radiation present it is not 
obvious that the surfaces of constant $\phi$ are still comoving with 
the matter. However this is indeed the case for adiabatic
perturbations as the following calculation
reveals.

We perform the calculation in conformal Newtonian gauge.
The equation governing the evolution of the radiation 
perturbation is given in this gauge  by
\be
(\delta_r -4 \Phi)' = {4\over 3} k v_r,
\labeq{radeq}
\ee
where $\delta_r$ is the fractional energy density perturbation in the
radiation and $v_r$ the velocity perturbation. We now make the assumption of 
adiabaticity, equivalent to the statement that the ratio of radiation
to scalar field energy density is spatially uniform being determined 
locally by the microphysics of brane collision. The condition for the 
`entropy perturbation' $\eta$, defined above, to be zero at early times is
 that $\delta_r = {2\over 3}\delta_\phi$, where $\delta_\phi$ is
 the fractional density perturbation in the 
scalar field. In Newtonian gauge we have
\be 
\delta_\phi = 2 \left({\delta \phi'\over \phi_0'}-\Phi\right).
\labeq{dep}
\ee
The final ingredient in calculating $v_r$ is the equation 
for perturbations in the massless scalar field,
\be
\delta \phi'' + 2 {\cal H} \delta \phi' = 4 \phi_0'\Phi'-k^2 \delta \phi. 
\labeq{scalaeom}
\ee
>From the adiabaticity condition and 
equations (\ref{eq:radeq}), (\ref{eq:dep}) and (\ref{eq:scalaeom}),
as well as the background equation $\phi_0''+2 {\cal H} \phi_0'=0$,
one determines
\be
v_r= - k {\delta \phi \over \phi_0'}.
\labeq{vrad}
\ee
This is the velocity of the radiation in Newtonian gauge. 
We wish to transform to the time-orthogonal 
gauge in which $\delta \phi$ is zero. The transformation required 
is a shift in conformal time by $T=-\delta \phi/\phi_0'$ accompanied
by a shift in spatial coordinates with a potential $L$ such that 
$L'=-kT$, to
ensure the absence of time-space components of the metric
(in the notation of  Section III in Bardeen's paper\cite{bardeen}).
The result of these transformations is that in the $\delta \phi=0$
gauge, the velocity potential for the radiation equals
$v_r +L'$, which is zero. Therefore, for adiabatic initial conditions,
the radiation fluid is actually comoving with the massless scalar field.
Therefore the comoving slices correspond to constant 
scalar field slices, both before and after the bounce.

So we may therefore focus on the evolution of the energy
density perturbation on comoving slices, $\epsilon_m$, and 
attempt to match it across the bounce at $\tau=0$. 
The equation of motion for $\epsilon_m$ is straightforwardly
derived from (\ref{eq:neweeq}) by eliminating $v_s^{(0)}$,
to obtain
\ba
&&\epsilon_m''+ f(\tau)
\epsilon_m' +
g(\tau) \epsilon_m
=-
k^2 w\eta,
\labeq{perteq}
\ea
where 
\ba
&& f(\tau)={\cal H} (1+3 c_s^2-6w),\cr
&&  g(\tau)=
{\cal H}^2(9 c_s^2 +{9\over 2} w^2 -12 w -{3\over 2}) + k^2 c_s^2.
\ea

As an aside, we now infer the behavior of the
variable $\zeta$ at small $\tau$. One has
\ba
\zeta =-
{(a^2 {\cal H} \epsilon_m
)'
\over (1+w) k^{2} a^{2}}.
\labeq{zetanew}
 \ea
Employing the background relations
\ba
{\cal H}'=-{1\over 2}(1+3w) {\cal H}^2,\qquad w'=-3{\cal H} (c_s^2-w)(1+w),
\labeq{relat}
\ea
one obtains
\ba
\zeta' =  {{\cal H} c_s^2
 \epsilon_m \over 1+w}.
\labeq{zetaeq}
\ea
Since 
$\zeta'$ is down by $k^2$ relative to $\zeta$,
the 
rate of change of $\zeta$ is typically
much smaller than ${\cal H} \zeta$ for the low $k$ modes of interest.
So one might imagine $\zeta$ would be a good matching variable.
However, as noted above,
$\epsilon_m$ is finite as $\tau$ tends to zero. From Eq.~(\ref{eq:zetaeq})
and the fact that ${\cal H} \propto \tau^{-1}$ for small $\tau$,
one observes that $\zeta$ diverges logarithmically at small $\tau$.
Hence, as stated above,
in linear perturbation theory $\zeta$ is not a good variable for
establishing a matching condition at $\tau=0$.
Nevertheless, $\zeta$ does have some utility.
If one
separates out the short wavelength piece which is divergent,
the remaining long wavelength component of $\zeta$ is  finite and 
nearly constant.  This finite piece is then very useful, since it
gives the amplitude
of the final growing mode linear density perturbation in the
expanding Universe.

We now turn to an analysis of Eq. (\ref{eq:perteq}).
The first thing to note is that in the setup considered here,
 ${\cal H}$,
$w$ and $c_s^2$ all have expansions in terms of simple powers of $\tau$,
for either $\tau<0$ or $\tau>0$:
\ba
&&{\cal H} = {1\over 2} \tau^{-1} + h_0 +h_1 \tau +\dots,\cr
&&w = 1 +w_1 \tau +w_2\tau^2 +\dots,\cr
&&c_s^2 = 1 +c_1 \tau +c_2\tau^2 +\dots.
\labeq{expa}
\ea
This follows from computing 
\ba
{\cal H}&=& {1\over 2} {{a^2}'\over a^2}, \cr
w&=&{{1\over 2} \phi'^2 -a^2V
 \over {1\over 2} \phi'^2 +a^2V}
= {3 M_{Pl}^2 Q^2 -a^2 U \over 3M_{Pl}^2 Q^2 +a^2 U}, \cr
c_s^2&=& 1+{2\over 3} {V_{,\phi} a^5 \over a' \phi' a^2}
= 1+{2\over 3}  {a_1 U_{,0} +a_0 U_{,1} \over {\cal H} M_{Pl} Q},
\labeq{occont}
\ea
where $a^2= {1\over 4} (a_0^2-a_1^2)$, $U=Va^4$ and 
$Q\equiv {1\over 4}(a_0 a_1' -a_1 a_0')$. As discussed in
Ref. 4, the  variables 
$a_0=  2 a \,{\rm cosh}(\phi/\sqrt{6}M_{Pl})$,  and
$a_1= -2 a \,{\rm sinh}(\phi/\sqrt{6}M_{Pl})$ are
both finite at $\tau=0$. 
These variables
were introduced in Ref.~\ref{kost}, with motivation from the
five dimensional geometry. In the AdS case, $a_0$ and
$a_1$ are the scale factors on the positive and negative tension
boundary branes respectively. Although the four dimensional
Einstein frame scale factor $a$ tends to zero at an outer-boundary 
collision, $\phi$ tends to minus infinity in just such a manner as 
to leave
$a_0$ and $a_1$ finite at collision. 

The physics of bounce and 
reversal are discussed in Ref.~\ref{nonsing}, the results of which
are 
assumed here.
The key point for our analysis is that
the equations
of motion for $a_0$ and $a_1$,
\be
{3\over 2}  M_{Pl}^2 a_0''= {\partial U\over \partial a_0}
\qquad {3\over 2}  M_{Pl}^2 a_1''= - {\partial U \over \partial a_1},
\labeq{apmeqs}
\ee
are actually regular at 
$\tau=0$. It follows that 
all the variables in Eq.~(\ref{eq:occont}) 
have simple power series expansions
around $\tau=0$, as shown in Eq. (\ref{eq:expa}),
for either positive or negative $\tau$.

It will be important below to include a component of
radiation after the bounce. As discussed in Ref.~\ref{nonsing},
the density and temperature of radiation on the branes is actually finite at 
$a=0$, as long as the radiation couples to the scale factors
$a_0$ or $a_1$, which are both finite (and equal) 
as $\phi \rightarrow -\infty$. 
 If, for example,  $U$ is zero after the collision
but there is a density $\bar{\rho}_r$ of radiation produced on
the brane with scale factor $\bar{a}$ (in the simplest models,
we have $\bar{a}=a_0$ or $a_1$ for the positive or negative tension
branes respectively). Then
the expression for $w$ in Eq. (\ref{eq:occont}) is
replaced by
\ba
w&=&
{3 M_{Pl}^2 Q^2 +{1\over 3} (\bar{\rho}_r \bar{a}^4
) a^2
\over 3 M_{Pl}^2 Q^2 + (\bar{\rho}_r \bar{a}^4) a^2}.
\labeq{occontrad}
\ea

Before proceeding
to analyze Eq. (\ref{eq:perteq}),
it is important to recognize that the
expansion coefficients in Eq.~(\ref{eq:expa}) are not independent,
but are related by 
Eq.~(\ref{eq:relat}). These imply, for example,
that $h_0=-{3\over 16} w_1$ and $c_1={2\over 3} w_1$. 
The coefficient functions in Eq.~(\ref{eq:perteq}) have the
following expansions
\be
f(\tau)  = -{1\over \tau} +f_0+f_1 \tau +\dots, \qquad 
g(\tau) = g_{-1} \tau^{-1} +g_0 +g_1 \tau+\dots, 
\labeq{expafg}
\ee
with $f_0=-{13\over 8} w_1$,  $g_{-1}={3\over 4} w_1$,
and $g_0={21\over 32} w_1^2 + k^2$.
Now the general solution of the perturbation equation
(Eq.~(\ref{eq:perteq}))
is a sum of the two linearly independent
solutions
\ba
\epsilon_m&=& \epsilon_0 D(\tau) +\epsilon_2 E(\tau),
\labeq{sol}
\ea
where $\epsilon_0$ and $\epsilon_2$ are arbitrary constants, and
\ba
D(\tau) &=& 1 +d_1 \tau+ 
\hat d_2 \tau^2 {\rm ln} |\tau| +\hat d_3 \tau^3 {\rm ln} |\tau|
+d_3 \tau^3 + \dots,\cr
E(\tau) &=& 
\tau^2 +e_3 \tau^3 + \dots,
\ea
The equation of motion determines all the coefficients 
$d_n$, $n=1,3,4,\dots$,  $\hat d_n$, $n=2,3,\dots$ and $e_n$,
$n=3,4,\dots$. 
One finds 
$d_1=g_{-1}={3\over 4} w_1$, $\hat{d}_2= -{1\over 2} 
(d_1(f_0+g_{-1}) + g_0) = -{1\over 2} k^2$, and so on.
Since the coefficients $w_1$, $w_2$, $\dots$ change
across $\tau=0$, the two series expansions are different
for $\tau<0$ and $\tau>0$. We denote the coefficients 
in (\ref{eq:sol}), for $\tau<0$ and $\tau>0$ respectively, as
$\epsilon_0(0^-)$, $\epsilon_2(0^-)$, and
$\epsilon_0(0^+)$, $\epsilon_2(0^+)$.
The matching rule we seek should determine the latter
two constants in terms of the former.

It seems clear that we should match the amplitude of the finite
perturbation variable $\epsilon_m$ 
across $\tau=0$, and this fixes
$\epsilon_0(0^+)=\epsilon_0(0^-)$. But how should we determine
$\epsilon_2(0^+)$? The simplest prescription is just to
set  $\epsilon_2(0^+)=
\epsilon_2(0^-)$. This amounts to 
matching the amplitude of the linearly independent 
solution which vanishes at $\tau=0$, as well
as that which is finite at $\tau=0$.
This prescription is invariant
under redefining the independent solutions, e.g.
by adding an arbitrary amount of the solution 
$E(\tau)$ to $D(\tau)$.  
Matching any other non-singular
perturbation variable, defined
to be an arbitrary linear combination of $\epsilon_m$ and
$\epsilon_m'$ with coefficients which
are non-singular background variables (defined to
possess power series expansions in $\tau$, as above) will, with
the same prescription of matching the amplitudes of both linearly independent
solutions, also yield
precisely the same result.

This prescription is simple, but it is certainly not
unique, and we emphasize that a 
proper understanding of the correct matching
condition must
ultimately rest on a
better understanding of the singularity, either
directly from string or M theory or from a well defined
regularization
and renormalization procedure.

Note that any such matching rule applied at $\tau=0$ cannot 
match both $\epsilon_m$ and 
$\epsilon_m'$, as  would be appropriate at
a regular point of the differential equation. This is because 
$\epsilon_m$ and $\epsilon_m'$ {\it cannot} be independently
specified at $\tau=0$, just because $\tau=0$  is a singular
point of the differential equation. 
The value of 
$\epsilon_2$ 
is fixed by the limit as $\tau$ tends to zero of 
${1\over 2} (\epsilon_m(\tau)- \epsilon_0 D(\tau))''$,
so one needs to know $\epsilon_m''$ as $\tau$ tends to
zero, in order to 
determine $\epsilon_2$.

Recall that since $\zeta$ diverges logarithmically 
as $\tau \rightarrow 0$, it is not a good matching variable.
However, this divergence only affects the short wavelength 
part of $\zeta$, and the long wavelength part is still very
useful since it yields the amplitude of the growing mode
perturbations in the expanding phase. As we now see, 
the above prescription for matching $\epsilon_m$ actually
implies that 
the long wavelength part of $\zeta$ has a jump across $\tau=0$.
Using Eq.~(\ref{eq:relat}), we re-express Eq.~(\ref{eq:zetanew}) as
\ba
\zeta = - {{\cal H} \over (1+w) k^{2}} \left(\epsilon_m'
+{3\over 2} {\cal H} (1-w) \epsilon_m\right).
\labeq{zetajump}
 \ea
Substituting the above expansions, one finds the leading 
order behavior
\be
\zeta\sim {\epsilon_0\over 4} ({\rm ln} |\tau| -{1\over 2}) 
+k^{-2}({1\over 2} \epsilon_2 -{3\over 16} \epsilon_0 w^{(2)}),
\labeq{zetaj}
\ee
where $w^{(2)}\equiv w_2+{3\over 8} w_1^2$,
plus terms which vanish as $\tau$ tends to zero. The first term
is logarithmically divergent at $\tau=0$. 
However, since it is down by a factor
of $k^2$,
it rapidly becomes irrelevant as $\tau$ increases away from zero. 
The second term, the long wavelength piece $\zeta^{lw}$,
is the quantity we are actually interested in. This constant, 
long wavelength
piece is accurately conserved after the bounce as long as
the matter evolution remains adiabatic, and yields the amplitude
of the growing mode adiabatic density perturbation in the 
late Universe. 

As we have discussed above, $\epsilon_m$ is finite at $\tau=0$
and from Section III and IV, there
is no long wavelength contribution to $\zeta$
in the collapsing phase. It follows that  $\epsilon_2=
{3\over 8} 
\epsilon_0 w^{(2)}$ in that phase. One situation of sepcial 
interest is 
where $w^{(2)}=0$ in the collapsing
phase, where the potential is irrelevant at $\phi \rightarrow
-\infty$ and there is no radiation in the incoming state.
In this case, 
$\epsilon_2(0^-)=0$. In this case, we would obtain the
same final result from any 
matching rule which set 
$\epsilon_2(0^+)= A \epsilon_2(0^-)$, with any constant $A$.

The key point is
that generically {\it $w^{(2)}$ jumps } across $\tau=0$,
since the background equation of state changes at the
brane collision.
Matching $\epsilon_0$ and 
$\epsilon_2$ we obtain a long wavelength contribution to
$\zeta$ in the expanding phase,
\be
\zeta^{lw} (\tau>0) \sim -
{3\over 16} k^{-2} \epsilon_0 (w^{(2)<}-w^{(2)>}),
\labeq{zetajexp}
\ee
where $w^{(2)<}$ and $w^{(2)>}$ are the values of $w^{(2)}$ 
for $\tau<0$ and $\tau>0$ respectively. 
Since $\epsilon_m \propto k^2 \Phi$ and, as discussed 
earlier, $\Phi$ has a nearly scale invariant power spectrum,
it follows from 
Eq.~(\ref{eq:zetajexp}) that if $w^{(2)}$ undergoes a jump, then
$\zeta$ will 
inherit a scale invariant long wavelength piece. This is
our main result. In the following sections we will
study a simple example of a situation where
$w^{(2)}$ is discontinuous.

We shall need explicit formulae for the jump in 
$w^{(2)}$, and for $\epsilon_0$.
The formulae for $w^{(2)}$ 
are obtained from Eqs. (\ref{eq:occont})
and (\ref{eq:occontrad}) above. If we assume that
prior to $\tau=0$ there is no radiation, so that in addition
to scalar kinetic energy we have only the
potential $U$, then 
taking the limit as $\tau$ tends to zero from below 
one finds that
\be
w^{(2)<}= {2 U_0^2+{4\over 3} M_{Pl}^2 Q U_1 \over M_{Pl}^4 Q^2}, \qquad
U_1= U'|_0 = a_0'{\partial U \over \partial a_0}
+a_1'{\partial U \over \partial a_1},
\labeq{befo}
\ee
where 
$Q={1\over 4}(a_0 a_1'-a_1 a_0')$.
Likewise, for the expanding phase, 
if we assume $U=0$ 
but radiation is now present,
we obtain
\ba
w^{(2)>}= {8\over 27} {(\bar{\rho_r} \bar{a}^4)^2\over M_{Pl}^4 Q^2}. 
\labeq{afte}
\ea
Since in general
$w^{(2)<}\neq w^{(2)>}$, we infer that generically,
a scale invariant spectrum of perturbations will, 
with our prescription above,  
propagate across $\tau=0$ into the expanding hot big bang phase.

Finally, to compute the perturbation amplitude given in Eq. 
(\ref{eq:zetajexp})
we need $\epsilon_0$. This can be read off
 from the expression 
Eq.~(\ref{eq:phisol}) above, by translating the $\tau$ dependence
into $a'/a^3$ and employing the fact that the latter gives the
exact dependence for the long wavelength modes of interest
even when the potential breaks away from
the pure exponential form used in the first half of this paper. 
We find that at $\tau=0$,
\be
\epsilon_0= 
{4 k^{1-\nu} \over
3 M_{Pl} \sqrt{p}a_0 (a_0'-a_1')}.
\labeq{epsres}
\ee

To summarize the results of this section,
we have elaborated the conditions under which 
a scale invariant 
spectrum of perturbations survives the passage through
$a=0$. Basically this requires that the equation of
state, as parametrized by $w(\tau)$, have a discontinuous 
first or second 
derivative with respect to $\tau$ 
at $\tau=0$. This condition would appear to be quite generically
fulfilled, in any situation where entropy is generated 
at a brane collision. A key assumption in the calculation 
was that
the energy density
is dominated by scalar field kinetic energy as we approach 
$a=0$. This assumption is natural in the ekpyrotic 
scenario, where the scalar field represents the separation
of the boundary branes. 
We shall explore one particular, simplified
model in the next section.

\section{Background Evolution in a Nearly Light-like Bounce}

In this section we review and extend the description of
reversal from contraction to expansion, as elaborated in
Reference 4.  As discussed there, the background
evolution is described by the 
variables $a_0$ and $a_1$, which for the simplest brane model
(i.e. branes in AdS) represent the scale factors on the
positive and negative tension boundary branes. 
The four dimensional effective scale factor $a$
is given by ${1\over 2} \sqrt{a_0^2-a_1^2}$, and this vanishes at the bounce.

If the potential $V(\phi)$ vanishes as 
$\phi$ runs off to $-\infty$
(more precisely, if the quantity $a^4 V(\phi)$ vanishes),
then the Friedmann constraint equation implies that 
trajectory in the $(a_0,a_1)$-plane 
intersects the boundary of moduli space $a_0=a_1$
along a light-like direction \cite{nonsing}\null.
Then, if no radiation is produced on the branes, 
the trajectory simply reverses, corresponding to the matching condition
${a}_{0,1}'({\rm out}) = -{a}_{0,1}'({\rm in})$.
We describe this as an {\it elastic} collision,
since the internal states of the two branes are
unchanged by the collision.

However, at any finite velocity, 
the boundary brane collision must result in the production of
radiation 
on the branes, since it is a non-adiabatic 
process.  In the M theory context 
the corresponding string theory is weakly coupled near the collision, 
and this production of radiation should be computable once the 
correct matching conditions are understood.

Let us consider the case where the incoming state has no
radiation, and the potential $V(\phi)$ vanishes at $t=0$ and
thereafter. This requires that the potential
is turned off at collision, requiring a sudden and permanent 
change in the internal state of the branes. Associated
with this change, we assume that 
a small amount of 
radiation, with density $\bar{\rho_r}$ is generated, on a brane
with scale factor $\bar{a}$. The collision is therefore
inelastic. We parameterize the inelasticity
as follows. The Friedmann constraint  after collision reads
\begin{equation}
{a}_0'({\rm out})^2 -{a}_1'({\rm out})^2 =  {4 (\bar{\rho_r} 
\bar{a}^4) \over 3 M_{Pl}^2}.
\label{eq:fc}
\end{equation}
Since right hand side is positive, the
outgoing trajectory must be time-like in the $(a_{0},a_1)$-plane.
Since radiation redshifts as $\bar{a}^{-4}$, 
the expression $(\bar{\rho_r} \bar{a}^4 )$ is 
a constant. If radiation is generated on both branes, this term can 
be taken to represent the sum of the corresponding terms for 
both branes. 

We then define the 
the efficiency $\xi$ with which radiation is
produced by
\begin{equation}
\xi \equiv \frac{4 (\bar{\rho_r} \bar{a}^4)}{ 3 M_{Pl}^2 \,  {{a}_1'({\rm in})}^2},
\label{eq:efficiency}
\end{equation}
which with (\ref{eq:fc}) yields a single equation
for the two velocities
${a}_0'({\rm out})$ and ${a}_1'({\rm out})$. We need
another equation to fix both.

In the special case of dimensional reduction from five to four
dimensions,
there is a natural candidate for an approximately conserved
quantity, analogous to the total 
momentum for an inelastic particle collision. As mentioned
above, at small brane separations one has the
standard Kaluza-Klein result that the size of the 
extra dimension is proportional to 
 $ e^{\sqrt{2/3} \phi/M_{Pl}}$.
As stated above, we are assuming that
the potential $V(\phi)$ vanishes as $\phi$ tends to $-\infty$.
Other terms in the Lagrangian describing matter on the branes
may in principle
acquire $\phi$-dependence upon dimensional reduction. However,
the terms describing massless gauge fields and
fermions do not obtain any such $\phi$-dependence due to their
conformal invariance in four dimensions. (To see this, note that the 
four dimensional Einstein-frame metric $g_{\mu \nu}$ is 
conformally related to the four dimensional components of the five dimensional
metric.)
Therefore at the classical level, in the $\phi \rightarrow
-\infty$ limit, the Lagrangian describing gravity, $\phi$ and
four dimensional radiation possesses a global symmetry 
$\phi \rightarrow
\phi \, + \, {\rm constant}$, and it is plausible that 
the corresponding Noether
charge $Q$,
\begin{equation}
Q \equiv {1\over 4} (a_0 {a}_1' - a_1 {a}_0') \propto a^2 {\phi' },
\end{equation}
tends to a constant
as the collision approaches.
In this limit the
classical equation of motion of $\phi$ 
is just $Q'=0$. 

However, the sign of $Q$ must flip at the bounce. As 
argued in Ref. 4, this is essential in order that the
trajectory remains in the physical region 
of the $(a_0,a_1)$-plane. The reversal of $Q$
may be also understood by the following higher-dimensional argument.
The value of $Q$ at collision is proportional to
the time derivative of 
log$(a_0/a_1)$. 
The latter quantity is 
the distance, rather than 
the vector displacement, between the boundary branes.
Hence,
if the branes  come together and then draw apart, $Q$ must change sign
although its magnitude remains constant.
It is therefore natural to impose the $Z_2$ symmetry 
at $t=0$,
$Q \rightarrow -Q$, which should become
exact in the limit that the collision velocity approaches zero.

These arguments suggest that we parameterize 
$Q$-violation at the bounce (brane collision) using 
\begin{equation}
\Delta \equiv {Q({\rm out })+Q({\rm in}) \over Q({\rm in})},
\label{eq:chidef}
\end{equation}
where $\Delta$ is expected to be small. Equations
(\ref{eq:fc}), (\ref{eq:efficiency}) and
(\ref{eq:chidef}) together uniquely parameterize the final values
of $a_0'$ and $a_1'$ after  the bounce:
\begin{eqnarray}
{a}_0'({\rm out})&=& \left(1-\Delta +{\xi\over 4(1-\Delta)}\right) {a}_1'({\rm in})
 \cr
{a}_1'({\rm out})&=& -\left(1-\Delta-{\xi\over 4(1-\Delta)}\right) {a}_1'({\rm in}).
\label{eq:nullb}
\end{eqnarray}

An example of an ekpyrotic two-brane collision is shown 
in Figure 1, for a specific choice of the inter-brane 
potential. Both branes initially expand under the influence
of the attractive potential. But when they get close, 
and the potential rises to zero, this decelerates $a_0$
so that it begins to contract. At collision,
$a_0'({\rm in}) =-a_1'({\rm in}) $. Immediately
afterwards, $a_0'$
is positive and, for small $\xi$, $a_1'$ is negative. 

\begin{figure}
\begin{center}
{\par\centering \resizebox*{6in}{3in}{\includegraphics{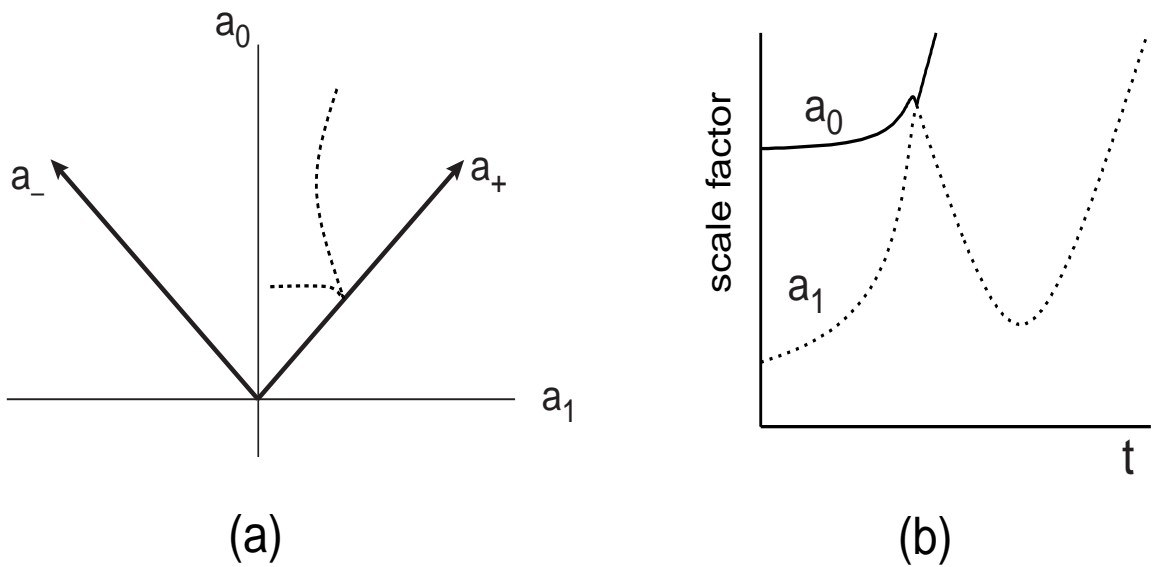}} \par}
\end{center}
\caption{Sketch of a nearly light-like collision 
between two boundary branes.
The potential employed was $a^4V(\phi) = -a_1^4 (a_1/a_0)^4 e^{-f}$ 
where $f= {1\over 10}\left((a_0/a_1)-1\right)^{-1}$, 
chosen so that when expressed
in terms of $\phi$, the potential $V$ vanishes 
as $\phi \rightarrow -\infty$, in a manner mimicking the
vanishing of a non-perturbative potential $V \propto e^{-{1\over g^2}}$
or $V \propto e^{-{1\over g}}$. 
At collision, $a_0'$ and $a_1'$ are 
equal and opposite. The matching rule we propose is given in
equation (\ref{eq:nullb});
in the Figure it is assumed that the efficiency $\xi$ and
$Q$-violation parameter $\Delta$ 
are
small. The scale factor $a_1$ is decreasing after collision:
we assume it couples to a massless modulus  $\chi$ which 
causes $a_1$ to be repelled from $a_1=0$. In the final state both
$a_0$ and $a_1$ are expanding, and the
universe becomes radiation-dominated with the 
outer-brane separation tending to 
a finite constant. 
 }
\label{fig:nullbounce}
\end{figure}

Assuming that the potential remains zero after collision,
and that only radiation is present, we have $a_0$ and
$a_1$ 
flying apart linearly in $t$ after collision, corresponding
to an expanding $a$.
According to (\ref{eq:nullb}), $a_1$ is  {\it decreasing} after 
collision, and it tends  to zero. This would lead to the
separation between the branes going to infinity.  However, it
is easily avoided, essentially because the $a_1$ modulus
has a {\it positive} kinetic term.
If there is one or more massless moduli fields $\chi$ 
coupling
to $a_1$ as $a_1^2 {\chi'}^2$, they produce an effective potential
in the $a_1$ equation which is proportional to $a_1^{-2}$. This 
repels the 
$a_1$ modulus 
from $a_1=0$.
Figure 1 shows an example of 
the full evolution, including the brane collision, reversal and
turn-around of $a_1$ so that $a_0$ and $a_1$ are both expanding at late times.
The final evolution of $a_1$ is 
insensitive to the value of $\chi'$, which only affects the
evolution at small $a_1$.
In the long time limit, the scalar field $\phi$ also tends to a constant
and therefore so does the inter-brane separation.

One
can generalize these considerations
to examples where the potential $U\equiv a^4 V(\phi)$ 
is finite and negative at brane collision.
This requires rather special potentials
$V(\phi)$, which diverge at large $\phi$, but not 
too strongly. In this case,
the trajectories are space-like at collision.

Having specified the background 
evolution for the `nearly light-like bounce', we can now
consider the matching of perturbations. We have assumed that
$U$ vanishes at collision. For simplicity we shall assume
that its first derivatives $\partial U /\partial a_0$ and $\partial U/
\partial a_1$ also
vanish there. Now we can read off from Eq. (\ref{eq:befo}) that
$w^{(2)<}=0$, but we have from Eq. (\ref{eq:afte}) and the formulae of
this section that
\be
w^{(2)>}= {2\over 3} \xi^2 {{a_1'({\rm in})}^2\over a_1^2}.
\labeq{nbr}
\ee
We can also read off from Eq. (\ref{eq:epsres}) that for the
`nearly light-like bounce', 
\be
\epsilon_0= 
-{2 k^{1-\nu} \over
3M_{Pl} \sqrt{p} a_1 a_1'({\rm in})},
\labeq{epsresn}
\ee
where we assumed $\xi, \Delta <<1$.
Putting these together in (\ref{eq:zetajexp}), we find the final density
perturbation amplitude is
\be
\zeta^{lw} \sim {1\over 12} \xi^2 {  k^{-1-\nu} \over
 M_{Pl} \sqrt{p}} 
{a_1'({\rm in})\over a_1^3}.
\labeq{finalamp}
\ee
It remains to compute $a_1$ and $a_1'({\rm in})$ at collision 
using the detailed 
behavior of 
the potential $V(\phi)$ for negative $\phi$.


As discussed in Ref. 4, as $\phi \rightarrow -\infty$,
the string coupling constant tends to zero. It is natural
to expect that the potential $V(\phi)$ 
goes to zero in this limit.
We shall adopt a very simplified model here, in which the 
potential jumps to zero at some particular,
negative  value of $\phi_j$.
In this case it is straightforward to compute
$a_1$ and $a_1'$ as  $\tau$ tends to zero from below.
First, assume the jump happens at some time $t_j$,
before which equations (\ref{eq:quants}) are valid.
The total energy in the scalar field is ${1\over 2} \dot{\phi}^2 
+V(\phi) = 3 p^2 M_{Pl}^2 /t^2 $, and this is equal to
the kinetic energy in the scalar field 
after the potential jump. So,
just after the jump we have
\be
\dot{\phi}= -\sqrt{6}  {p M_{Pl}\over (-t_j)}.
\labeq{velaft}
\ee
Now, from $a_0\equiv 2 a {\rm cosh} (\phi/\sqrt{6} M_{Pl}),$ we have
\be
a_0'= a\left (Ha_0-{\dot{\phi}\over \sqrt{6} M_{Pl}} a_1\right)
\labeq{avels}
\ee
and similarly for $a_1'$. Since the potential vanishes,
and no radiation is present, $a_0'$ and $a_1'$ are both
constant
up to collision. Then equations (\ref{eq:back}),
(\ref{eq:quants}) and (\ref{eq:avels}) 
imply that 
\be
a_0'({\rm in})=-2 p (-t_j)^{2p-1} e^{\phi_j/(\sqrt{6} M_{Pl})}= -a_1'({\rm in}).
\labeq{velsj}
\ee
After the potential jump, $a_0$ and $a_1$ evolve linearly
in $\tau$. To leading order in $\xi$ and $\Delta$ we have 
\ba
a_0&=& (-t_j)^p 
\left(2 {\rm cosh}(\phi_j/\sqrt{6} M_{Pl})+a_1'({\rm in})(\tau-\tau_j)\right)\cr
a_1&=& (-t_j)^p 
\left(2 {\rm sinh}(-\phi_j/\sqrt{6} M_{Pl})-a_1'({\rm in})(\tau-\tau_j)\right).
\labeq{linear}
\ea
Setting these equal determines the time of collision and brane scale factors at
collision,
\be
a_0=a_1=  (-t_j)^p e^{-\phi_j/(\sqrt{6} M_{Pl})}.
\labeq{branesc}
\ee
Now we have all we need to determine the final
fluctuation spectrum. From equations (\ref{eq:finalamp}), (\ref{eq:velsj})
and (\ref{eq:branesc}), we have
\be
\zeta^{lw} \sim {1\over 6} {\sqrt{p}
 \over
 M_{Pl}
  k^{1+\nu} (-t_j)^{1+p} }
\xi^2 e^{4\phi_j/(\sqrt{6} M_{Pl})}.
\labeq{finalampf}
\ee
The dependence on $k$, 
$t_j$ and $p$ is
the same as that one obtains from  the naive `time delay'
formula mentioned in Section II.
Using equation 
(\ref{eq:dps}) for $\delta \phi$, the time delay method
yields a perturbation amplitude $\sim H \delta \phi/\dot{\phi}
 \sim M_{Pl}^{-1} \sqrt{p} k^{-(1+\nu)}
(-t)^{-(1+p)}$, in agreement with the dependence upon these quantities 
in equation
(\ref{eq:finalampf}). In the time delay argument, however, one
uses the Hubble constant on the branes at collision, 
which is close to but 
not quite the same as the factor occurring in (\ref{eq:finalampf}).

Let us now translate the dependence
of the last factor in equation (\ref{eq:finalampf}) into 
quantities determined by observations in the 
final expanding, radiation dominated Universe.
After collision, we assume the potential $V(\phi)$ is zero 
(because the internal state of the branes has changed so they
no longer attract each other), so that
the $\phi$ modulus describing the inter-brane separation is a free
massless field. We also assume radiation is present, in an 
abundance parameterized by $\xi$.  
It is straightforward to analytically solve the equations of motion
post-collision, assuming the presence of  
the additional modulus needed to keep $a_1$ away from zero. This modulus
becomes irrelevant at late times. 
In the long time limit, one finds $a_0$ and $a_1$ increasing linearly
in conformal time $\tau$, with
$a_0/a_1 \rightarrow |a_0'({\rm out})/a_1'({\rm out})|
\approx 1+{1\over 2} \xi$, neglecting the dependence on
$\Delta$ which is reasonable if $\Delta$  is small.
Equating this to coth$(-\phi/\sqrt{6} M_{Pl})$,
we find that the {\it final} resting value for $\phi$
is given by
\be
e^{(2 \phi_f / \sqrt{6} M_{Pl})} = {\xi \over 4}.
\labeq{finalphi}
\ee
However, what enters equation (\ref{eq:finalampf}) is 
not $\phi_f$, but $\phi_j$, the value of $\phi$ 
at which the potential $V(\phi)$ switches off.
We can translate both values of $\phi$ into
the corresponding string coupling constants, 
using the relation $g_s\propto e^{\sqrt{3/2} \phi/M_{Pl}}$,
which 
follows from M theory with the assumption that 
the six Calabi-Yau dimensions
are fixed in the 11 dimensional metric \cite{wittengut}\null.

For $p$ close to zero, we find the final
result 
\be
k^{3\over 2} \zeta^{lw}_k  \sim \sqrt{-V_j \over M_{Pl}^4} 
{\xi^4 \over 96 } \left({g_s(j) \over g_s(f)}\right)^{4\over 3}.
\labeq{finalampfin}
\ee
The right hand side is the amplitude of the growing mode
adiabatic density perturbation relevant to structure formation in 
the late Universe. (For an accurate calculation 
one should of course retain the $p$ dependence, since over the 
many orders of magnitude of $k$ involved, this can significantly
affect the final normalization. We leave this complication 
for future work.)

Our result for $\zeta^{lw}_k $ depends 
on the square root of the 
potential energy $V$ at its minimum, in Planck units. This is 
reminiscent of the usual inflationary result. However, additional
suppression factors arise. First, the numerical coefficient
is small. Second, the factor $\xi^4$ is small if the 
efficiency of production of radiation at collision is small. 
Finally, the  string coupling constant
where the potential turns off, which we have crudely parameterized
as $g_s(j)$, would be expected to be substantially
smaller than the value of the string coupling constant 
in the asymptotic outgoing state.
Translating the formulae relevant to 
Horava-Witten theory, 
given by Witten\cite{wittengut}, one finds for today's value 
of the string coupling constant
\be 
g_s(f)= \left({M_{Pl}\over M_{GUT}}\right)^3 {g_{GUT}^4\over 
\sqrt{2}}v_0^{1\over 2} ,
\labeq{gsf}
\ee
where the volume of the Calabi-Yau manifold is $v_0 M_{GUT}^{-6}.$

Before discussing numerical values, it is important to make the following
caveats. First, the resting value of the scalar field $\phi$ determined
during the radiation era following the bounce is not necessarily that 
measured in today's universe. If there is a stabilizing potential for
$\phi$, that will instead determine the final resting value. Nevertheless
it is conceivable that the resting value early in the hot big bang phase
is closely related to the final value (as for example if $\phi$ develops
a potential with many closely spaced degenerate minima). Second, 
the presence of additional moduli (such as are found in 
Horava-Witten theory) could have important consequences 
on the dynamical evolution of $\phi$ in these early stages. In the above
calculation we have limited ourselves to only one modulus, translating
that directly into the string coupling constant. So the final numerical
result can only be suggestive. 

For example, plausible values of the GUT coupling are
$g_{GUT}^2 \sim 0.5$, and the GUT mass $10^{17}$ GeV, giving 
$g_s(f) \sim 10^3 v_0^{1\over 2}$. The 
turn-off of non-perturbative potentials might plausibly occur at
$g_s(j) \sim 10 $, if instanton effects produce factors
of the form
exp($-8 \pi^2 /g^2$). The
last factor in (\ref{eq:finalampfin}) then
yields $\sim 10^{-5}v_0^{-{2\over 3}}$. For $v_0 \sim 10^{-3}$, and 
$\xi^4 (-V_j )/M_{Pl}^4\sim 10^{-2}$, we can obtain 
an amplitude $\sim 10^{-5}$, as required
by observations.

We conclude that
the ekpyrotic scenario may offer a natural explanation for
the smallness of the observed 
 density perturbations. As we have emphasized, this is
only suggestive at this stage, and will remain so in the absence
of:

$\bullet$ a microscopic check of the matching condition
used for $\epsilon_m$, within the context of M theory
and string theory,

$\bullet$ a computation of the efficiency parameter $\xi$
describing the production of radiation on the branes,

$\bullet$ a check that the parameter $\Delta $ is indeed
small, as was assumed, 

$\bullet$ and, a full calculation of a
realistic inter-brane potential $V(\phi)$ and a numerical
solution of the equations improving the jump approximation
used above. 

Although we have focused here on matching scalar
perturbations, in principle one also needs a matching
condition for tensor and vector modes. Neither acquire long wavelength
power in the contracting phase of the ekpyrotic model, and it seems
unlikely they will be generated at the brane collision. Nevertheless
one can attempt to study possible matching conditions. It is not hard
to see that the tensor amplitude $h_{ij}^T$ exhibits the same logarithmic
divergence as the perturbation in the three-curvature of comoving
slices, $\zeta$. However, the canonically conjugate momentum
$a^2 {h_{ij}^T}'$ does tend to a finite constant at $\tau=0$, suggesting
it provides a possible matching variable. Again, establishing this
will probably require a satisfactory microscopic theory.


\section{Conclusions}

We have shown that inclusion of gravitational backreaction
has a negligible effect on the density perturbations
produced in the ekpyrotic universe, during the initial phase
which is slowly contracting from the point-of-view of the 
four dimensional effective theory.  

We have proposed  what we believe is
a physically 
sensible matching condition at the bounce, 
based upon identifying physical 
perturbation variables which are well behaved (i.e. small) 
at $\tau=0$, and matching on surfaces defined by the scalar field
$\phi$.
The variables we used are well-behaved, we should emphasize, provided 
certain conditions are met as the 4d effective scale factor 
approaches zero, namely (1)~the energy density is dominated by 
the kinetic energy of a scalar field (the modulus $\phi$ in our
case) and (2)~the inter-brane potential $V(\phi)$ does not
diverge as $\phi$ tends to $-\infty$.
We showed that with the simplest matching prescription,
namely matching the two linearly independent solutions
at $\tau=0$, the scale invariant spectrum of perturbations 
developed in $\Phi$ early on in the contracting phase is generically
passed on to the variable $\zeta$ representing the amplitude of
the long wavelength growing mode density perturbation in the 
expanding phase. We also identified examples where no 
density perturbations are generated in the final Universe.
If no radiation is generated at the brane collision, and
if the potential vanishes sufficiently smoothly there,
the perturbations `time-reverse' at $a=0$ and the amplitude of the
growing mode perturbation is precisely zero in the final expanding 
Universe. Such examples are not realistic since there is 
no entropy generation at the
brane collision, and in any case seem highly unlikely given
that the outer-brane collision is not an
adiabatic event.

 The existence of a `zero-perturbation limit' is an intriguing
feature of the ekpyrotic model and the matching prescription given here,
since it suggests  a natural
explanation for the smallness of the observed density perturbations.
Recall that a feature of inflation is that it naturally predicts
a value of the density perturbation amplitude that is far too large,
and fine-tuning of potentials is required to obtain a sufficiently
small amplitude.

We conclude that, while many issues connected to the microphysics
at brane collision remain to be settled by rigorous investigation of
string theory in the limit of outer-brane collision,
the basic idea of 
Ref.~\ref{kost} for producing density perturbations during the
early stages of the ekpyrotic Universe, in a phase which
is slowly contracting from the four dimensional point of view,
remains viable. 
\bigskip

{\bf Acknowledgements:} 
We thank V. Mukhanov for insightful remarks and contributions.
We have greatly benefited from explaining our arguments and criticisms
to D. Lyth prior to publication and obtaining his comments.
We especially thank D. Wands for helpful correspondence and
important questions. 
We also thank R. Brandenberger for showing us a preliminary version
of his preprint.
This work was supported in part by
 the Natural Sciences and Engineering Research Council of
Canada (JK),
the  US Department of Energy grants
DE-FG02-91ER40671 (JK and PJS) and  DE-AC02-76-03071 (BAO), and
by PPARC-UK (NT).

{\bf Note Added:}
As mentioned in the Introduction, our conclusions 
differ  from those of Lyth, Brandenberger and
Finelli, and Hwang\cite{lyth2,brand2,hwang}.
The disagreement is 
due to different assumptions
about physical conditions near the bounce, and to 
a different prescription for matching across it.
We believe that the assumptions made
by the authors of Refs. 7-9 are 
inconsistent with what we proposed in Refs. 1 and 4, because

$\bullet$ they use a scalar field potential which diverges to minus 
infinity as the bounce is approached. This potential is not
compatible with the bounce prescription discussed in Ref. 4,
since it is too singular. 

$\bullet$ they choose to match on a surface of constant energy density
in the contracting phase, assuming that the scale factor of the universe
reverses from contraction to expansion on this surface. This behavior
is incompatible with the field equations describing the cosmological
background solution.
We instead follow the classical field equations all the way to
the bounce, and apply a matching prescription consistent with
our treatment of the background at this point.
As we have explained above
the appropriate 
matching surface in the ekpyrotic setup is defined by the scalar
field specifying the inter-brane separation, and not by the energy density.

$\bullet$ Brandenberger and Finelli claim our matching 
prescription based on the energy density perturbation on slices of
constant scalar field is inconsistent with results given in the literature 
for models where the equation of
state undergoes a sudden jump\cite{deru}\null. It is easy to see why
matching $\epsilon_m$ and its time derivative in this situation is incorrect.
The equation of motion (\ref{eq:perteq})
 for $\epsilon_m$ involves $c_s^2$, related by (\ref{eq:relat})
to the time derivative of $w$. Hence, if $w$ jumps, $c_s^2$ acquires 
a delta function contribution, which causes a jump in 
$\epsilon_m'$ across the matching point, which may be straightforwardly
computed. The key point, however, is in the
situation we are discussing, $w$ {\it is continuous} across the bounce, 
thus there are no such delta function contributions.
Therefore this `counterexample' is truly a red herring.

$\bullet$ Brandenberger and Finelli mistakenly imply
that reversal from contraction to expansion would occur
at a bulk brane-boundary brane collision.
If the four dimensional effective description is valid, 
then reversal can {\it only}
happen at a boundary brane-boundary brane collision, as described in Ref. 4.
In the four dimensional effective description 
the universe continues to contract after a bulk-boundary 
collision, so 
$\zeta$ remains small and $\Phi$ continues to grow. Only when
the outer boundary brane collision  occurs as it must,
can the growing perturbations
in $\Phi$ get converted to long wavelength fluctuations in $\zeta$.


\begin{thebibliography}{9999}
\bibitem{kost}  J. Khoury, B.A. Ovrut,
P.J. Steinhardt and N. Turok, hep-th/0103239.
\label{kost}
 \bibitem{GuthPi} 
  A. Guth and S.-Y. Pi, {\it Phys. Rev. Lett.} {\bf 49}, 1110 (1982).
  \bibitem{Wang} L. Wang, V. Mukhanov and P.J. Steinhardt,
  {\it Phys. Lett. B}{\bf 414}, 18 (1997).
\bibitem{nonsing}
 J. Khoury, B.A. Ovrut,
 N. Seiberg,
 P.J. Steinhardt and N. Turok,
hep-th/0108187.
 \label{nonsing}
\bibitem{lyth0} D. Lyth, hep-ph/0106153.
\label{lyth0}
\bibitem{lyth1}
D.H. Lyth and E.D. Stewart, {\it Phys. Lett B}{\bf 274}, 168 (1992).
\bibitem{lyth2} D. Lyth, Lancaster preprint, in preparation (2001).
\label{lyth2}
\bibitem{brand2} R. Brandenberger and F. Finelli, hep-th/0109001.
\label{branden}
\bibitem{hwang} J. Hwang, astro-ph/0109045.
\bibitem{bgt} M. Bucher, A.S. Goldhaber and N. Turok,
{\it Phys. Rev. D}{\bf 52},  3314 (1995).
\label{bgt}
\bibitem{bardeen} J.M. Bardeen, Physical Review {\bf D22} (1980)
1882. 
\bibitem{Mukh}
V. F. Mukhanov, {\it JETP Lett.} {\bf 41}, 493 (1985);
Sov. Phys. JETP
{\bf 68} (1988) 1297;
 V.F. Mukhanov, H.A. Feldman and R.H. Brandenberger, Phys. Reports
{\bf 215} (1992) 203. 
\bibitem{BST} J.M. Bardeen, P.J. Steinhardt and M.S. Turner, 
Phys. Rev. D {\bf 
28} (1983) 679.
\bibitem{Brand} R. Brandenberger and  R. Kahn, 
{\it Phys. Rev. D}{\bf 29},
2172 (1984).
\bibitem{gt} 
S. Gratton and N. Turok, Physical Review {\bf D60} (1999) 123507.
\label{gt}
\bibitem{wittengut} E. Witten, hep-th/9602070.
\bibitem{deru} 
N. DeRuelle and V.F. Mukhanov, gr-qc/9503050, Phys.Rev. D52 (1995) 5549.
\label{deru}
\end{thebibliography}
\end{document}